\documentclass{article}

\usepackage{microtype}
\usepackage{graphicx}
\usepackage{booktabs} 

\usepackage{hyperref}

 \usepackage[accepted]{mlsys2022}

\mlsystitlerunning{QuClassi: A Hybrid Deep Neural Network Architecture based on Quantum State Fidelity}

\newcommand{\sol}{\texttt{QuClassi}}
\usepackage{xcolor}
\usepackage{graphicx}              
\usepackage{caption}
\usepackage{subcaption}
\usepackage{algorithm,multirow}
\usepackage{algorithmic}
\usepackage{amsmath}
\usepackage{comment}

\begin{document}

\twocolumn[
\mlsystitle{QuClassi: A Hybrid Deep Neural Network Architecture based on Quantum State Fidelity}




\begin{mlsysauthorlist}
\mlsysauthor{Samuel A. Stein}{1,2}
\mlsysauthor{Betis Baheri}{3}
\mlsysauthor{Daniel Chen}{4}\\
\mlsysauthor{Ying Mao}{1}
\mlsysauthor{Qiang Guan}{3}
\mlsysauthor{Shuai Xu}{4}
\mlsysauthor{Caiwen Ding}{5}
\mlsysauthor{Ang Li}{2}
\end{mlsysauthorlist}

\mlsysaffiliation{2}{Pacific Northwest
National Laboratory, Richland, WA, USA}

\mlsysaffiliation{1}{Department of Computer and Information Science, Fordham University, New York, USA}
\mlsysaffiliation{3}{Department of Computer Science, Kent State University, Kent, OH, USA}

\mlsysaffiliation{4}{Department of Computer and Data Sciences,  Case Western Reserve University, Cleveland, OH, USA}

\mlsysaffiliation{5}{Department of Computer Science \& Engineering, University of Connecticut, Storrs, CT, USA}

\mlsyscorrespondingauthor{Samuel A. Stein}{samuel.stein@pnnl.gov}
\mlsyscorrespondingauthor{Ying Mao}{ymao41@fordham.edu}

\mlsyskeywords{Machine Learning, MLSys}

\vskip 0.3in

\begin{abstract}
In the past decade, remarkable progress has been achieved in deep learning related systems and applications.  
In the post Moore’s Law era, however, the limit of semiconductor fabrication technology along with the increasing data size has slowed down the development of learning algorithms.
In parallel, the rapid development of quantum computing has pushed it into a new era. Google illustrated quantum supremacy by completing a specific task (random sampling problem), in 200 seconds, which continues to be impracticable for the largest classical computers. Due to the exponential potential of quantum computing,  
quantum based learning is an area of interest, in hopes that certain systems might offer a quantum speedup. 
In this work, we propose a novel architecture \sol, a quantum neural network for both binary and multi-class classification. Powered by a quantum differentiation function along with a hybrid quantum-classic design, \sol~encodes the data with a reduced number of qubits and generates the quantum circuit, pushing it to the quantum platform for the best states, iteratively. We conduct intensive experiments on both quantum simulators, IBM-Q's quantum platform as well as evaluate performance on IonQ which was accessed through Microsoft's Azure Quantum Platform. The evaluation results demonstrate 
that \sol~ is able to outperform the state-of-the-art quantum-based solutions, Tensorflow-Quantum and QuantumFlow by up to 53.75\% and 203.00\% for binary and multi-class classifications. When comparing to traditional deep neural networks, \sol~ achieves a comparable performance with 97.37\% fewer parameters. 
\end{abstract}
]



\printAffiliationsAndNotice{}  

\section{Introduction}\label{sec:introduction}

%
%
%
%


Deep learning (DL) has drawn tremendous interest in the past decade from industry and academia~\cite{goodfellow2016deep, devlin-etal-2019-bert,lecun2015deep,he2016deep}.
Novel DL algorithms, growing computational power and modern architectural designs have enabled a wide spectrum of applications ranging from scientific data processing\cite{CosmoGan, Wozniak2018} 
to image and speech recognition \cite{NIPS2012_4824,6296526,amodei2016deep}. Despite the widespread practical success, the proximity to the physical bound of semiconductor fabrication in post Moore's Law era along with the increasing size of data sets raises the discussion on the future of DL and its limitations~\cite{waldrop2016chips}. 

In parallel with the breakthrough of DL in the past years, remarkable progress has been achieved in the field of quantum computing. In 2019, Google demonstrated Quantum Supremacy using  a 53-qubit quantum computer, where it spent 200 seconds to complete a random sampling task that would cost 10,000 years on the largest classical computer~\cite{supremacy}. During this time, quantum computing has become increasingly available to the public. IBM Q Experience, launched in 2016, offers 
quantum developers to experience the state-of-the-art superconducting quantum computers~\cite{ibmq}. Early 2020, Amazon Braket~\cite{aws} provides the general access to a development environment to help customers explore and design quantum algorithms. 
Microsoft Quantum Development Kit (QDK) \cite{QDK} bridges the gap between quantum computing and classic computing, and provides private access to trapped-ion machines by IonQ and Honeywell.

Quantum computing is a computational paradigm that harnesses quantum phenomena and the computational advantages that quantum computers offer~\cite{quantumcomputing}. A traditional computer with von Neumann architecture completes tasks by manipulating bits that can be set to 0 or 1. Everything from tweets and emails to electronic tax returns and healthcare records are essentially long strings of these binary digits.
A quantum computer, however, utilizes quantum bits, or qubits, to encode information. Qubits exhibit quantum properties, which means a connected group of them have exponentially more expressive and processing power than the same number of binary bits. 
Two fundamental properties that enable this phenomena are superposition and entanglement. 

A qubit, similar to bits, has two basis states $|0\rangle$ and $|1\rangle$. However, the value of quantum computing comes from the fact that a qubit can be in a superposition of $|0\rangle$ and $|1\rangle$ at the same time. 
As for entanglement, entangled qubits can be generated by specific interactions within a quantum system. Qubits that are entangled can not be described independently and only as a single quantum state. For a specific pair, changing the state of either qubit will affect the state of the other qubit predictably. This property in quantum physics is known as entanglement. 

Due to the great potential of processing complex problems beyond current abilities at a fraction of the time, quantum based learning systems have received great attention recently. 
The authors in~\cite{schuld2015introduction} provide a systematic overview of the emerging field of quantum machine learning. Many  researchers  try  to find  quantum  algorithms  that  can  take  the  place of  classical  machine learning  algorithms, and show an improvement in terms of a computational complexity  reduction \cite{PhysRevLett.109.050505,PhysRevA.99.052331}. 
Widely used classic machine learning algorithms such as nearest neighbor, the kernel method and other clustering algorithms, which are comprised of expensive distance calculations have been proposed to be accelerated by the design of a quantum counterpart~\cite{qnn2015}.
Recent innovations~\cite{aimeur2007quantum, rebentrost2014quantum, casana2020probabilistic, chen2020quantum, xue2021quantum, chen2021end, li2021quantum, stein2021hybrid, stein2021qugan, chen2020quantuminspire}
in the field of quantum machine learning for classification mainly focus on solving binary classification problems. 
With deep quantum neural network architectures and data pooling, many solutions proposed act and perform very similarly to their classical counterparts.  Although useful, the binary classification setting significantly limits the scope of the solutions.
Furthermore, many of the proposed approaches are still suffering from relatively low accuracy, missing a comprehensive quantum architecture that performs multi-class classification, or requiring an infeasible number of qubits for data encoding and high parameter counts.

In this work, we propose \sol, an architecture for high-count multi-class classification with a limited number of qubits. Making use of a quantum state fidelity cost function, \sol~utilises a data encoding architecture encoding two dimensions of data per one qubit. \sol~provides leading classification accuracies within the quantum DL field on the MNIST~\cite{mnist} and Iris~\cite{iris} datasets. Moreover, 
\sol~has been evaluated with binary and multi-class experiments on both local clusters, IBM-Q~\cite{ibmq} (Superconducting Quantum Processor (QPU)) and IonQ (Trapped Ion QPU). We summarize the key contributions below.

\begin{itemize}
    \item  We introduce a quantum state fidelity based cost function that enables the stable training of deep quantum circuits for classification.
    
    \item We propose \sol, a quantum-classical architecture with three different layer designs. 
    The \sol~works for both binary and multi-class classification problems.
    
    \item We evaluate \sol~with well-known datasets, performing a complete both binary and multi-class classification. \sol~ is evaluated with experiments on both the simulator as well as a real quantum platform, IBM-Q. When comparing with Tensorflow, Tensorflow-Quantum and the state-of-the-art QuantumFlow~\cite{jiang2020}, we achieve accuracy improvements by up to $53.75\%$ on binary classification and $203.00\%$ on multi-class classification. Comparing with classical deep neural networks with similar performance, \sol~ is able to reduce 97.37\% of the parameters.

\end{itemize}

The rest of this paper is organized as follows. 
In Section~\ref{sec:background}, we introduce the background of quantum computing and related techniques that are utilized in this work. 
We present the system design of \sol~ in Section~\ref{sec:design} and, in Section~\ref{sec:results}, we discuss the results from intensive experiments on local quantum simulator environment and IBM-Q platform.
We conclude the paper discussion with Section~\ref{sec:conclusion}. Acknowledgements (See Sec \ref{sec:acknowledgements}) and Appendix follow.




\section{Related Work}\label{sec:relatedwork}

With the recent advances in this field, quantum computing introduces exciting possibilities to enhance the existing learning algorithms and architectures through qubits, the basic unit in quantum information theory. 
Great efforts have been made to develop a quantum based learning algorithms.
In~\cite{garg2020advances}, authors conduct a comparative study of classic DL architectures with various Quantum-based learning architecture from a different perspective. 
The first challenge researchers encountered is how to represent classical data (binary states) with  quantum states. Different methods have been proposed to address it~\cite{cortese2018loading,zoufal2019quantum,lloyd2018quantum}. Cortese et al.~\cite{cortese2018loading} discusses set  quantum circuits and associated techniques that are capable to efficiently transfer binary bits from the classical domain into the quantum world. However, the proposed methods require $Log_2(\mathcal{N})$ qubits and $O(Log(\mathcal{N}))$ depth to load $\mathcal{N}$ bits classical data. Aiming to reduce the complexity, qGANs~\cite{zoufal2019quantum} utilizes quantum Generative
Adversarial Networks to facilitate efficient learning and loading of generic probability distributions, which are implicitly are given by data samples, into quantum computers.

    
 As quantum computers quickly evolve, quantum neural networks quickly gained attention in the field. A neural network architecture for quantum data has been proposed in~\cite{farhi2018classification}, which
utilizes the predefined threshold to build a binary classifier for bit strings. 
Focusing on optimizing variational quantum circuits, Stokes et al.~\cite{stokes2020quantum} introduce a quantum generalization of natural gradient for general unitary and noise-free quantum circuits. 
A design of quantum convolutional neural network is proposed in ~\cite{cong2019quantum} 
with guaranteed $O(Nlog(N))$ variational parameters for input sizes of N qubits that enables efficient training and implementation on realistic, incoming quantum computers.
Based on a general unitary operator that acts on the corresponding input and output qubits, authors in~\cite{beer2020training} propose a training algorithm for this quantum neural network architecture that is efficient in the sense that it only depends on the width of the individual layers and not on the depth of the network.
More recently, QuantumFlow~\cite{jiang2020} proposes a co-design framework that consists of a quantum-friendly neural network and an automatic tool to generate the quantum circuit. However, it utilizes quantum computers by training the network locally and then mapped to quantum circuits. This design is easy to implement, but shows significant sensitivity to real quantum computer noise. In addition, it is still based on a loss function based on classical data.

The existing works primarily investigate binary classification, which is an important problem. This problem setting, however, significantly limits the scope of the designs. Furthermore, many solutions suffer from having low accuracies, missing a comprehensive quantum architecture, and requiring an infeasible number of qubits. 
We propose \sol~ that employs a quantum state fidelity based loss function and a quantum-classic hybrid architecture to address the current limitations.

\section{Quantum Computing Basics}\label{sec:background}


\subsection{Quantum Bits - Qubits}
Traditional computers are built on the premise of information being represented as either a 0 or a 1. In contrast, quantum computers represent data through the use of a 1, 0, or both 1 and 0 simultaneously. This ability to represent a 1 and 0 simultaneously is a quantum mechanical phenomena called superposition, that is at the core of quantum computers computational potential. These quantum bits, namely qubits, are the fundamental building block of quantum computers. The qubits are represented in the form of linear vector combinations of both ground states $|0\rangle$ and $1\rangle$. Quantum systems are identified through the $\langle bra||ket\rangle$ notation, where $\langle bra|$ is a horizontal state vector and $|ket\rangle$ is a vertical state vector. A qubit can be mathematically represented with Equation \ref{eq1}:

\begin{small}
\begin{equation}
\label{eq1}
|\psi\rangle=\alpha|0\rangle+\beta|1\rangle
\end{equation}
\end{small}

where $|\psi\rangle$ is the qubit itself, and $|0\rangle$ and $|1\rangle$ represent the "0" and "1" translation to quantum systems that the qubit is a linear combination of. The $|0\rangle$ and $|1\rangle$ represent two orthonormal eigen vectors, of which $|\psi\rangle$ is a probabilistic combination. These ground state vectors are described in Equation \ref{eq:state_vec}:

\begin{small}
\begin{equation}
|0\rangle=\left[\begin{array}{l}
1 \\
0
\end{array}\right],|1\rangle=\left[\begin{array}{l}
0 \\
1
\end{array}\right],|\Psi\rangle=\left[\begin{array}{l}
\alpha \\
\beta
\end{array}\right]
\label{eq:state_vec}
\end{equation}
\end{small}

With superposition, there are multiple new possibilities for data representation on quantum system, where instead of classical binary data encoding, data could be represented in the form of a qubits probability distribution, or expectation. Most importantly, the data representation architecture of quantum computers opens up a wide range of possible data encoding techniques. 
Taking the tensor product between the qubits described in Equations  \ref{eq:xtra_qubit} and \ref{eq1} in \ref{eq:tensor_example} describes the state of two qubits.

\begin{small}
\begin{equation}
|\phi\rangle=\gamma|0\rangle+\omega|1\rangle
\label{eq:xtra_qubit}
\end{equation}
\end{small}

\begin{small}
\begin{equation}
|\Psi \phi\rangle=|\Psi\rangle \otimes|\phi\rangle=\gamma \alpha|00\rangle+\omega \alpha|01\rangle +\gamma  \beta|10\rangle+\omega \beta|11\rangle
\label{eq:tensor_example}
\end{equation}
\end{small}

The coefficients to each respective quantum state described in Equation \ref{eq:tensor_example}, and all quantum state equations in general, describe the probability of measuring each respective outcome. Therefore, a constraint is introduced that the squared sum of the coefficients is equal to 1. A qubit, when measured, will measure as either of the $|0\rangle$ or $|1\rangle$. These probabilities indicate the likelihood of measuring a $|0\rangle$ or $|1\rangle$. The measurement of $|0\rangle$ and $|1\rangle$ is arbitrary, and is simply one "direction" of measurement. The only constraint on direction of measurement is that the outcome of the measurement is one of two orthogonal eigen vectors. The bloch sphere is a common way of representing a qubit, which can be thought of as a unit sphere, and the quantum state is a point on this sphere's surface. The $|0\rangle$ and $|1\rangle$ vectors indicate the outer sphere points on the Z axis. However, one could measure a qubit to be on the X axis, or the Y axis, or any axis described by two rotations, touching two opposite ended points on the unit sphere. For the purposes of this paper, we will only be measuring against the Z-axis resulting in measurements of $|0\rangle$ and $|1\rangle$.
Quantum states are responsible for encoding data, and to perform operations on quantum states quantum gates are used. Quantum gates apply a transformation over a quantum state into some new quantum state.

\subsection{ Quantum Gates and Operations}
\subsubsection{Single qubit operations}
A quantum computers controlling apparatus allows the application of a single-qubit rotation $R(\theta,\phi)$ to any single qubit by unitary operator described by 
Equation \ref{eq2}:

\begin{small}
\begin{equation}\label{eq2}
    R(\theta,\phi) = \left[ \begin{array}{cc} \cos{\frac{\theta}{2}} & -ie^{-i\phi}\sin{\frac{\theta}{2}} \\-ie^{i\phi}\sin{\frac{\theta}{2}} & \cos{\frac{\theta}{2}} \end{array} \right]
\end{equation}
\end{small}

Both $\theta$ and $\phi$ 
can be 
controlled by changing the duration and phase of the R that drives the Rabi oscillation of the qubit\cite{debnath2016demonstration}.

\subsubsection{RX, RY, RZ rotation}

The single-qubit rotations around the basis axes must be expressed in terms of general R rotation defined in Equation \ref{eq2}. 
We can separate each rotation
in respect to rotations 
around the \texttt{X,Y and Z} axis. 
\textbf{RX} can be defined by setting $\phi=0$ in $R(\theta,\phi)$ the rotation about the X axis by the angle $\theta$, as described in Equation \ref{eq3}:

\begin{small}
\begin{equation}\label{eq3}
    R_X(\theta) = \left[ \begin{array}{cc} \cos{\frac{\theta}{2}} & -i\sin{\frac{\theta}{2}} \\-i\sin{\frac{\theta}{2}} & \cos{\frac{\theta}{2}} \end{array} \right] = R(\theta,0)
\end{equation}
\end{small}

Similarly \textbf{RY} is defined by setting $ \phi = \frac{\pi}{2}$ in $R(\theta,\phi)$ obtaining the rotation about Y axis by the angle $\theta$ 
described 
in Equation \ref{eq4}:

\begin{small}
\begin{equation}\label{eq4}
    R_Y(\theta) = \left[ \begin{array}{cc} \cos{\frac{\theta}{2}} & -\sin{\frac{\theta}{2}} \\\sin{\frac{\theta}{2}} & \cos{\frac{\theta}{2}} \end{array} \right] = R(\theta,\frac{\pi}{2})
\end{equation}
\end{small}

Finally, \textbf{RZ} rotation is defined in Equation \ref{eq5}:

\begin{small}
\begin{equation}\label{eq5}
    R_Z(\theta) = \left[ \begin{array}{cc} e^{-i\frac{\theta}{2}} & 0 \\ 0 & e^{i\frac{\theta}{2}} \end{array} \right]
\end{equation}
\end{small}

\subsubsection{Two-qubit gates and Controlled Operation}
Quantum gates can operate on more than one qubit simultaneously. The basic Rotation operation forming a single qubit rotation can be expanded to \texttt{$R_{XX}$}, \texttt{$R_{YY}$} and \texttt{$R_{ZZ}$} rotations.
$R_{XX}(\theta)$ are defined in Equations \ref{eq9} - \ref{eq11}:

\begin{small}
\begin{gather}
\label{eq9}
    R_{XX}(\theta) = \left[ \begin{array}{cccc} \cos{\frac{\theta}{2}} & 0 & 0 & -i\sin{\frac{\theta}{2}} 
    \\ 0 & \cos{\frac{\theta}{2}} & -i\sin{\frac{\theta}{2}} & 0 
    \\ 0 & -i\sin{\frac{\theta}{2}} & \cos{\frac{\theta}{2}} & 0 
    \\ -i\sin{\frac{\theta}{2}} & 0 & 0 & \cos{\frac{\theta}{2}}\end{array} \right]
\end{gather}
\end{small}
\begin{small}
\begin{gather}
\label{eq10}
    R_{YY}(\theta) = \left[\begin{array}{cccc} \cos{\frac{\theta}{2}} & 0 & 0 & i\sin{\frac{\theta}{2}} 
    \\ 0 & \cos{\frac{\theta}{2}} & -i\sin{\frac{\theta}{2}} & 0 
    \\ 0 & -i\sin{\frac{\theta}{2}} & \cos{\frac{\theta}{2}} & 0 
    \\ i\sin{\frac{\theta}{2}} & 0 & 0 & \cos{\frac{\theta}{2}}\end{array} \right]
\end{gather}
\end{small}
\\
where \texttt{$R_{ZZ}(\theta)$} is defined such that
\begin{small}
\begin{gather}
\label{eq11}
    R_{ZZ}(\theta) = \left[`\begin{array}{cccc} e^{-i\frac{\theta}{2}} & 0 & 0 & 0 
    \\ 0 & e^{-i\frac{\theta}{2}} & 0 & 0 
    \\ 0 & 0 & e^{-i\frac{\theta}{2}} & 0 
    \\ 0 & 0 & 0 & e^{-i\frac{\theta}{2}} \end{array} \right]
\end{gather}
\end{small}

Another key operation in data manipulation is the Controlled-SWAP (CSWAP) gate, allowing for Quantum entanglement, which can be defined as: 
\begin{small}
\begin{gather*}
\label{eq8}
    CSWAP(q_0,q_1,q_2) = 
    |0\rangle\langle0| \otimes I \otimes I + |1\rangle \langle1| \otimes \mbox{SWAP}
	\\
	\\
    = \left[ \begin{array}{cccccccc} 1 & 0 & 0 & 0 & 0 & 0 & 0 & 0  
    \\ 0 & 1 & 0 & 0 & 0 & 0 & 0 & 0
    \\ 0 & 0 & 1 & 0 & 0 & 0 & 0 & 0
    \\ 0 & 0 & 0 & 0 & 0 & 1 & 0 & 0
    \\ 0 & 0 & 0 & 0 & 1 & 0 & 0 & 0
    \\ 0 & 0 & 0 & 1 & 0 & 0 & 0 & 0
    \\ 0 & 0 & 0 & 0 & 0 & 0 & 1 & 0
    \\ 0 & 0 & 0 & 0 & 0 & 0 & 0 & 1 \end{array} \right]
\end{gather*}
\end{small}
where $q_0$, $q_1$, $q_2$ are qubits entangled to each other. We further make use of Controlled rotations, which are extensions on the \texttt{RY,RZ and RX} gates however with a control operation appended to them.

\subsection{Quantum Entanglement}
Along with superposition, quantum entanglement is another phenomenon of quantum computing that exposes significant computing potential. Quantum entanglement is described as when two quantum states interact in a certain way,  following the interaction the quantum states can not be described independently of each other. This generally implies that the outcome of one qubits measurement implies information about the other. A common experiment is Bell's experiment, which is when two qubits that have been entangled through a CNOT gate, allows for the measurement of one qubit to, with absolute certainty, predict the measurement of the second qubit. As a concrete idea, if qubit $|\phi\rangle$, entangled with qubit $|\psi\rangle$, measures as a $|0\rangle$, $\psi\rangle$ will measure as $|0\rangle$, and vice versa. This principle of quantum entanglement occurs when a controlled gate is used, as described in Equation \ref{eq8}, and with gates such as controlled rotations, where a control qubit dictates an entangled change in state on the other qubits. 

A key use case of entanglement is the SWAP test, a quantum state fidelity measurement algorithm that operates off of quantum entanglement. The SWAP test is a probabilistic distance metric that measures the difference between two quantum states, supposedly $|\phi\rangle$ and $|\omega\rangle$. The SWAP test takes in the two states, performs a CSWAP over the states controlled by an anicilla qubit. This qubit, prior to the SWAP test, is placed in superposition by a Hadamard gate from state $|0\rangle$. After the CSWAP, we apply another Hadamard gate to the anicilla qubit, and measure the qubit. The probability of the qubit measuring 1 ranges from $0.5$ to $1$. If the states are orthogonal, the qubit will measure 1 approximately $50\%$ of the time, and as the quantum state fidelity increases as does the probability of measuring a $|1\rangle$ increases, eventually at $100\%$. 


\section{QuClassi: Quantum-based Classification}\label{sec:design}

\subsection{System Design}
\begin{figure}[ht]
\centering
\includegraphics[width=0.95\linewidth]{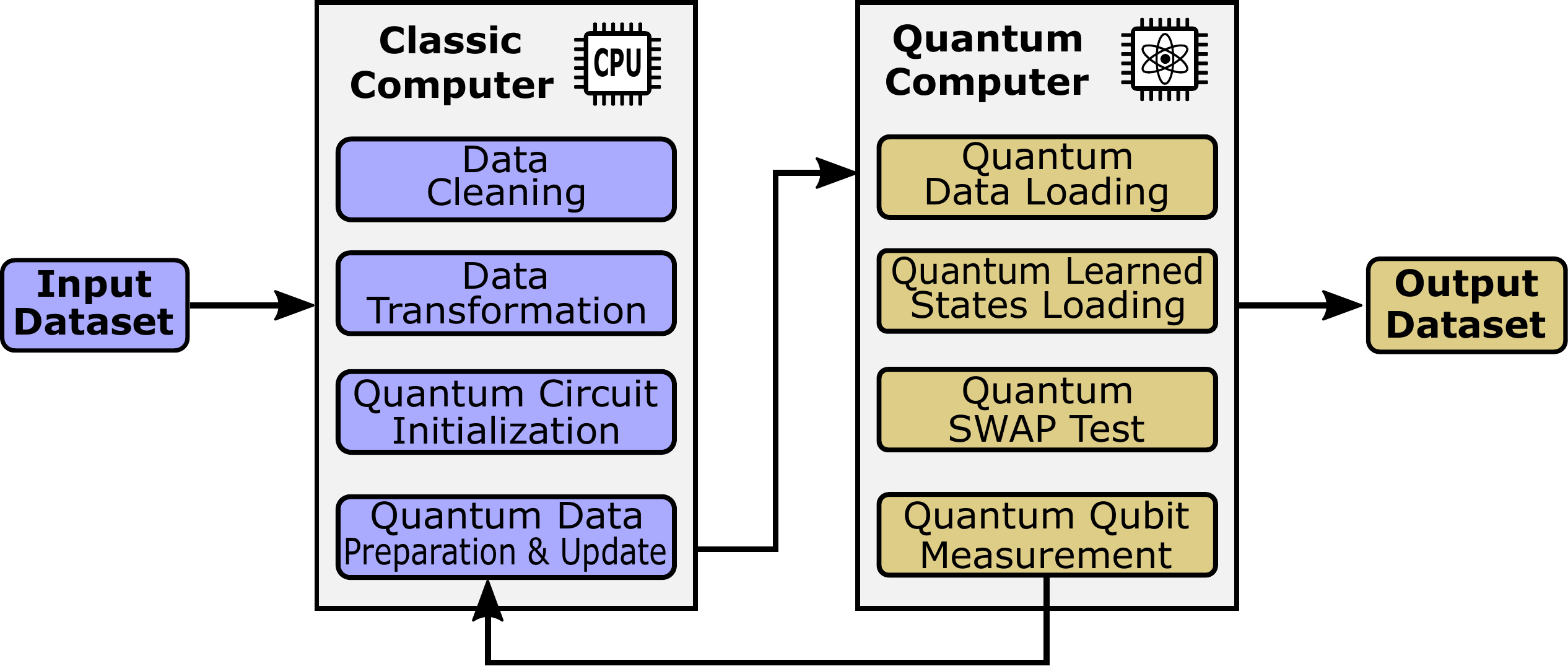}
\caption{Quantum-Classic Deep Learning.}
\label{fig:qmlsys}
\end{figure}

Our architecture operates through a feedback loop between a classical computer and a quantum computer, illustrated in Fig.~\ref{fig:qmlsys}. Data is initially cleaned, removing any significant outliers or any other necessary data cleaning steps. Data is parsed from classical data into quantum data through a quantum data encoding method outlined in 4.2, and is now in the form of a  quantum data set represented by quantum state preparation parameters. For each predictable class in the data set, a quantum state is initialized with the same qubit count as the number of qubits in the classical quantum data set, due to the constraints of the SWAP test. The quantum states along with quantum classical data are sent to a quantum computer.

This initialization of state is the core architecture to QuClassi. In this, a quantum circuit of a certain number of layers representing a quantum deep neural network is prepared with randomly initialized parameters containing a certain number of qubits. The produced quantum state of this circuit is to be SWAP tested against the quantum data point, which is fed back to the classical computer, forming the overall quantum DL architecture of QuClassi. 

The quantum computer calculates the quantum fidelity from one anicilla qubit which is used to calculate model loss, and sends this metric back to the classical computer. The classical computer uses this information to update the learn-able parameters in attempts to minimize the cost function. This procedure of loading quantum states, measuring state fidelity, updating states to minimize cost is iterated upon until the desired convergence or sufficient epochs have been completed.

\subsection{Data Qubitization}
\label{encode}
When evaluating quantum machine learning architectures on classical datasets, we need a way to translate classical data into quantum states. An important question is how one might represent a classical data set in the quantum setting. Our architecture makes use of translating some traditional numerical data point into the expectation of a qubit. To accomplish this, data ${x_1,x_2,...,x_n}$ of dimension ${d}$ can be translated onto a quantum setting by normalizing each dimension ${d_i}$ to be bound between ${0}$ and ${1}$ due to the range of a qubits expectation. Encoding a single dimension data point only requires one qubit, unlike classical computing which requires a string of bits to represent the same number. To translate the traditional value $x_i$ into some quantum state, we perform a rotation around the Y axis parameterized by the following equation: 
  $  RY(\theta_{x_i}) = 2sin^{-1}(\sqrt{x_i}) $.
This $RY(\theta_{x_i})$ operation results in the expectation of a qubit being measured against the Z axis equating to the $x_i$ value from the classical data the qubit encodes. An extension on this idea is that we can encode our second dimension of data across the X-Y plane. We make use of two parameterized rotation on one qubit initialized in state $|0\rangle$ to prepare classical data in the quantum setting. To encode a data point, we prepare the required rotations across ${\frac{d}{2}}$ qubits, each rotation parameterized by each dimension normalized value of that data point. It is worth noting that the 2-dimensional encoding onto one qubit can be problematic in extreme values of $x$, however we explore the dual dimensional encoding as a possible method of alleviating high qubit counts and evaluate the performance if we encode each dimension of data into one respective qubit solely through a RY Gate. This approach is validated by the fact that we never measure any of our qubits, and only their quantum fidelity through the SWAP test, thereby bypassing the superposition-collapsing issue of this approach. We encode the second dimension of data on the same qubit through the following rotation:

\begin{equation}
    RZ(\theta_{x_{i+1}}) = 2sin^{-1}(\sqrt{x_i})
\end{equation}

When number of qubits is a primary concern, methods that will reduce this number are extremely valuable. 
Classical data encoding in quantum states has no method that is tried and tested unlike classical computers with formats such as integers and floats, therefore our approach can certainly be critiqued; however the approach was tested and proved to be a viable approach to the problem. Furthermore, knowing both the expectation of a qubit across the Y and Z domain allows for the data classical data to be reconstructed. Alternative methods exist for classical-to-quantum data encoding, where on one extreme, one can encode $2^n$ classical data points across $n$ qubits using state-vector encoding. However, this method is extremely susceptible to noise. On the other end of the spectra, one can encode classical data into a binary representation on quantum states, by translating a vector of binary values onto qubits. This encodes $n$ data points onto $n$ qubits, and loses a lot of information in the process. However, is not as susceptible to noise and exponential-sampling problems. Exponential data-encoding methods exist however, and can be fitted with \sol due to \sol never directly performing quantum state tomography, hence the data encoding section is scalable.

\subsection{Quantum Layers}
The design of our quantum neural network can be broken down into a sequential layer architecture. Each layer proposed in this paper is designed with the intent of each layer having a specific type of quantum state manipulation available to it. We introduce three layer styles, namely single qubit unitaries, dual qubit unitaries, and controlled qubit unitary. Single qubit unitaries involve rotating a single qubit by a set of parameters ${\theta_d}$. 
\begin{figure}[!htb]
\centering
         \includegraphics[width=0.75\linewidth]{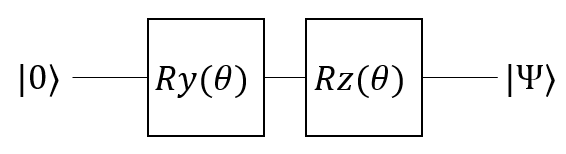}
\caption{Single Qubit Unitary.}
      \label{fig:squ}
\end{figure} 
A qubit enters a single qubit unitary, drawn in Fig.~\ref{fig:squ}, in a certain state and undergoes a rotation around the Y and Z axis. A rotation around the Z and Y axis allows for complete manipulation of a single qubits quantum state. 

\begin{figure}[ht]
\begin{minipage}{0.45\columnwidth}
\centering
\includegraphics[width=1\columnwidth]{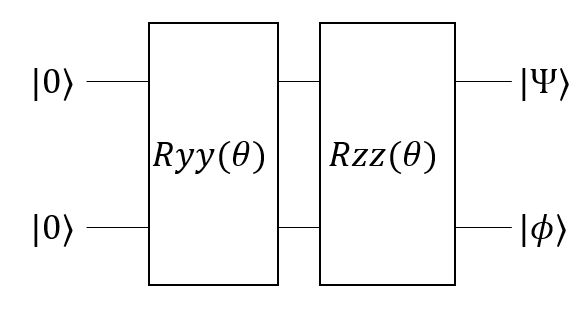}
\caption{Dual Qubit.}
      \label{fig:dqu}
\end{minipage}
\hfill
\begin{minipage}{0.45\columnwidth}
\centering
         \includegraphics[width=1\columnwidth]{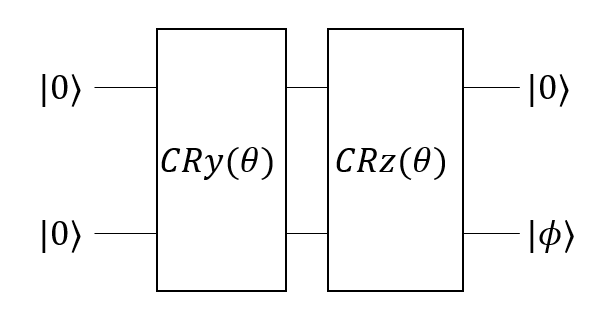}
\caption{Entanglement.}
      \label{fig:equ}
\end{minipage}      
\end{figure}

As for layers comprised of gates operating on more than one qubit simultaneously, we make use of the Dual Qubit unitary and Entanglement unitary. With a dual qubit unitary, drawn in Fig.\ref{fig:dqu}, two qubits enter in their each respective gate, followed by an equal Y and Z rotation on both qubits. The rotation performed on both qubits are equal. 
Entanglement layers, drawn in Fig.\ref{fig:equ}, involve two qubits being entangled through two controlled operations per qubit pair. We introduce the use of $CRY(\theta)$ and $CRZ(\theta)$ gates within this layer, thereby allowing for a learn-able level of entanglement between qubits. In the above layer design, two qubits are passed in with one qubit being a control qubit and the other the target qubit. With respect to a $CRY$ gate, if the control qubit measures $|1\rangle$, the operation of $RY(\theta)$ will have happened on the other qubit, similarly with the $CRz(\theta)$ operation.

\begin{figure}[ht]
\centering
         \includegraphics[width=0.75\linewidth]{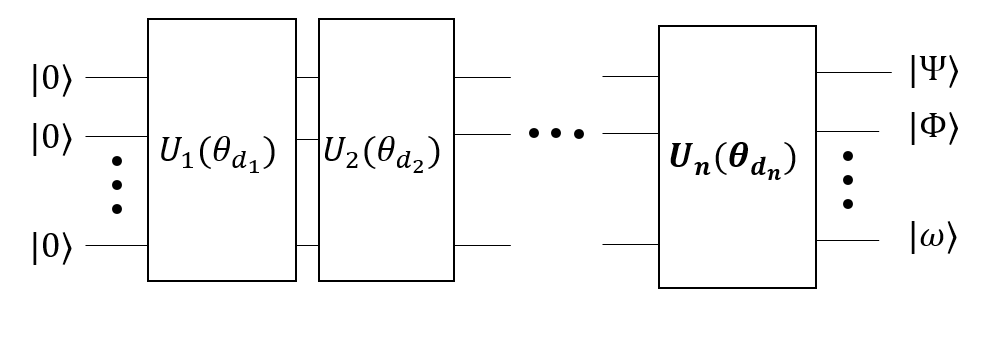}
\caption{Layered Quantum Deep Learning Architecture.}
      \label{fig:lqdla}
\end{figure}

Each of the aforementioned layers contains rotations parameterized by a group of parameters, which are the trainable parameters in our neural network. 
We can combine a group of each of the unitaries on a quantum state to form a quantum layer, parameterized by each independent $\theta$ in the layer, where the layer is set as a group of these operations, as visualized in Figure \ref{fig:lqdla}.

The use of these layers, drawn in Fig.~\ref{fig:lqdla}, allows for the abstraction to  the design that is similar to the traditional DL. Each layer is similarly parameterized by some set of values $\theta_d$. These parameters can be trained such that the final output of the quantum circuit minimizes some cost function.

\subsection{State fidelity based Cost Function}
When training a neural network to accomplish a task, an explicit description of system improvement goal needs to be established - i.e the cost function. The quantum machine learning cost function landscape can be slightly ambiguous compared to classical machine learning, as we could be manipulating the expected values of each qubit in some way, however even this is ambiguous - the direction being measured in heavily affects the expectation value and or  what our iteration count would be for measuring expectation, with lower iterations leading to increasingly noisy outputs. Within our system, we make use of the SWAP test to parse quantum state fidelity to an appropriate cost function. One of the benefits of the SWAP test is that we only need to measure one anicilla qubit.In the case of binary classification, each data point is represented in a quantum state represented by $|\phi\rangle$, which is used to train the quantum state prepared by our DL model $|\omega\rangle$ such that the state of $|\omega\rangle$ minimizes some cost function. The classical cross-entropy cost function outlined in Equation \ref{eq:cross_ent} is an appropriate measure for state fidelity, as we want the fidelity returned to be maximized in the case of Class=1, and minimized otherwise.
\begin{equation}
\label{eq:cost_func}
    min(Cost(\theta_d,X) = \frac{1}{n}\sum_{i=1}^{n} SWAP(|\phi_{X(i)}\rangle,|\omega\rangle)
\end{equation}
\begin{equation}
\label{eq:cross_ent}
    Cost = -y log(p) - (1-y) log(1-p)
\end{equation}
Where $\theta_d$ is a collection of parameters defining a circuit, $x$ is the data set, $\phi_{x(i)}$ is the quantum state representation of data point ${i}$, and ${\omega}$ is the state being trained to minimize the function in Equation \ref{eq:cost_func} and \ref{eq:cross_ent}. \\
Optimization of the parameters $\theta_d$ requires us to perform gradient descent on our cost function. We make use of the following modified parameterized quantum gate differentiation formula outlined in Equation \ref{eq:diff}.
\begin{equation}
    \label{eq:diff}
    \frac{\delta Cost}{\delta\theta_i} = \frac{1}{2}(f(\theta_i + \frac{\pi}{2\sqrt{\epsilon}}) - f(\theta_i - \frac{\pi}{2\sqrt{\epsilon}}))
\end{equation}

Where in Equation \ref{eq:diff} $\theta_i$ is a parameter, Cost is the cost function, and $\epsilon$ is the epoch number of training the circuit. Our addition of the $\epsilon$ is targeted at allowing for a change in search-breadth of the cost landscape, shrinking constantly ensuring a local-minima is found.

\subsection{Training and Inducing QuClassi}
\label{training}

We implement the Algorithm outlined as Algorithm 1 for the training of our Quantum DL architecture.

\begin{algorithm}[!t]
\caption{QuClassi Algorithm}
\label{alg:1}
\begin{algorithmic}[1]
\STATE Data set Loading 
    \par \quad $\text {Dataset}: (X | Class = mixed) $
\STATE Distribute Dataset X By Class
\STATE Parameter Initialization: 
		\par \quad $\text{Learning Rate}: \alpha  = 0.01$
		\par \quad $\text{Network Weights}: \theta_d = [\text{Rand Num btwn 0 - 1}\times\pi]$
		\par \quad $\text{epochs}: \epsilon= 25$
		\par \quad $\text {Dataset}: (X | Class = \omega) $
		\par \quad $\text {Qubit Channels}: Q = n_{X_{dim}}\times2$
\FOR {$\zeta \in \epsilon$}
    \FOR {$x_k \in X$}
        \FOR {$\theta \in \theta_d$}
            \STATE Perform Hadamard Gate on $Q_0$
            \STATE Load $x_k \xrightarrow[Data\text{ }Encoding]{Quantum} Q_{Q_1\xrightarrow[]{}\frac{Q_{Count}}{2}+1}$ 
            \STATE Load $\theta_d \xrightarrow[Learned\text{ }State]{Quantum} Q_{\frac{Q_{Count}}{2}+1\xrightarrow[]{}Q_{Count}}$ 
            \STATE Add $\frac{\pi}{2\sqrt{\epsilon}} \xrightarrow[]{} \theta$
            \STATE $\Delta_{fwd} =  (\mathbf{E}_{Q_0} f(\theta_d)$ 
            \STATE CSWAP(Control Qubit = $Q_0$, Learned State Qubit, Data Qubit)
            \STATE Measure $Q_0$
            \STATE Reset $Q_0$ to $|0\rangle$
            \STATE Perform Hadamard Gate on $Q_0$
            \STATE Subtract $\frac{\pi}{2\sqrt{\epsilon}} \xrightarrow[]{} \theta$      
            \STATE CSWAP(Control Qubit = $Q_0$, Learned State Qubit, Data Qubit)
            \STATE Measure $Q_0$
            \STATE $\Delta_{bck} =  (\mathbf{E}_{Q_0} f(\theta_d)$ 
            \STATE $\theta = \theta - (-log(\frac{1}{2}\Delta_{fwd}-\Delta_{bck})) \times\alpha $
        \ENDFOR
    \ENDFOR
\ENDFOR
\end{algorithmic}
\end{algorithm}

Algorithm \ref{alg:1} describes how \sol~works. Firstly, we load the dataset as presented in~\ref{encode} (Line 1). Line 2-3 introduce certain terms which are chosen by the practitioner. The learning rate is a variable chosen at run time that determines at what rate we want to update our learned weights at each iteration. Qubit Channels indicate the number of qubits that will be involved in the quantum system. Epochs indicate how many times we will train our quantum network on the dataset. 
Within the nested loops, Lines 7-23, we load our trained state alongside our data point with a forward difference applied to the respective $\theta$ ($\Delta_{fwd}$), perform the SWAP test, reset and perform the backward difference to aforementioned $\theta$ ($\Delta_{bck}$). These lines 7-23 accomplish the specific parameter tuning of our network, where $f(\theta_d)$ is the overall cost function of the network.
At induction time, the quantum network is induced across all trained classes and the fidelity is softmaxed. The highest probability returned is the classified class.

\section{Evaluation}\label{sec:results}

\begin{figure*}
\begin{subfigure}{.33\textwidth}
\centering
         \includegraphics[width=1.0\linewidth]{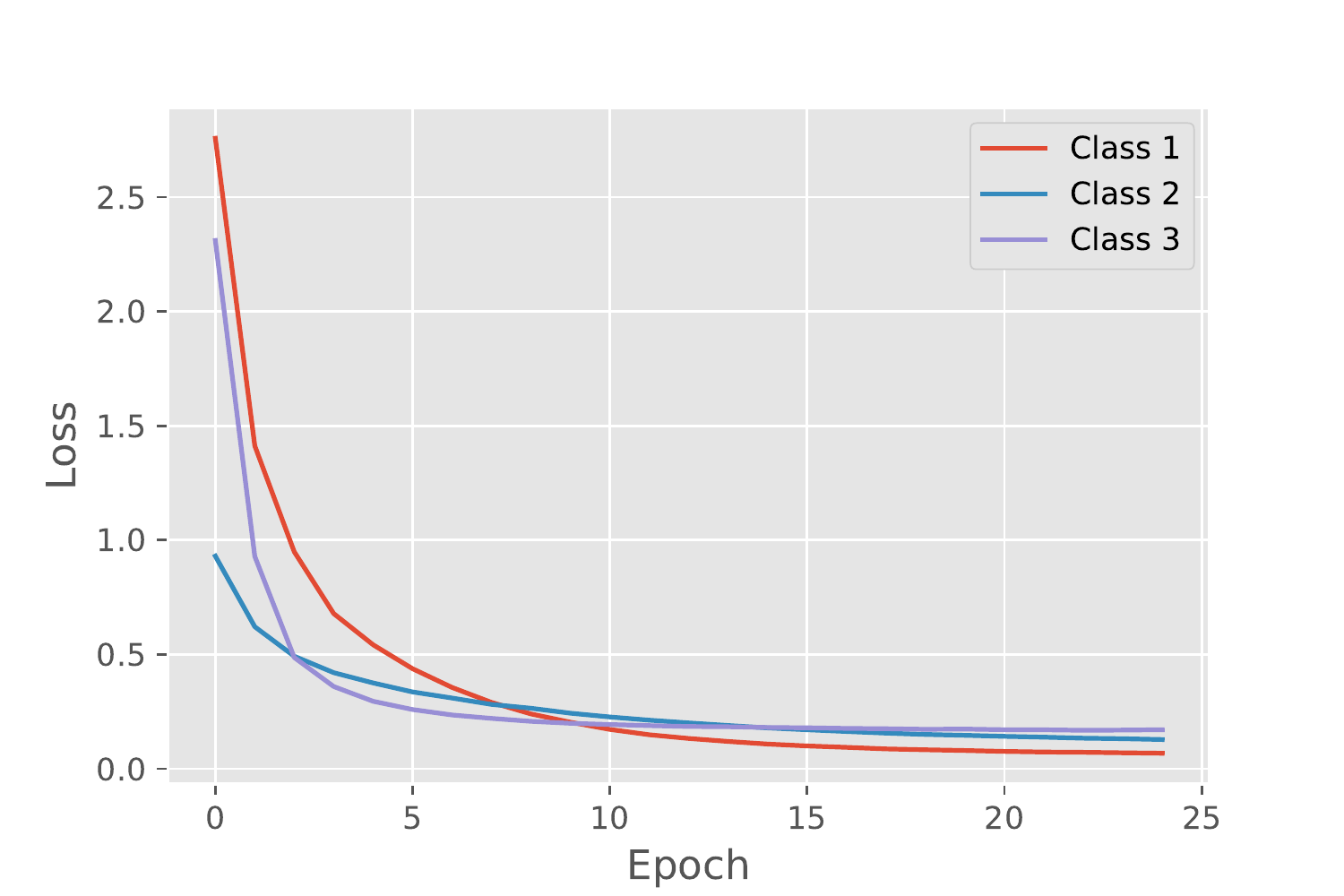}
\caption{Multi-class Loss Values.}
      \label{fig:multi_class}
\end{subfigure}
\begin{subfigure}{.32\textwidth}
\centering
         \includegraphics[width=1.0\linewidth]{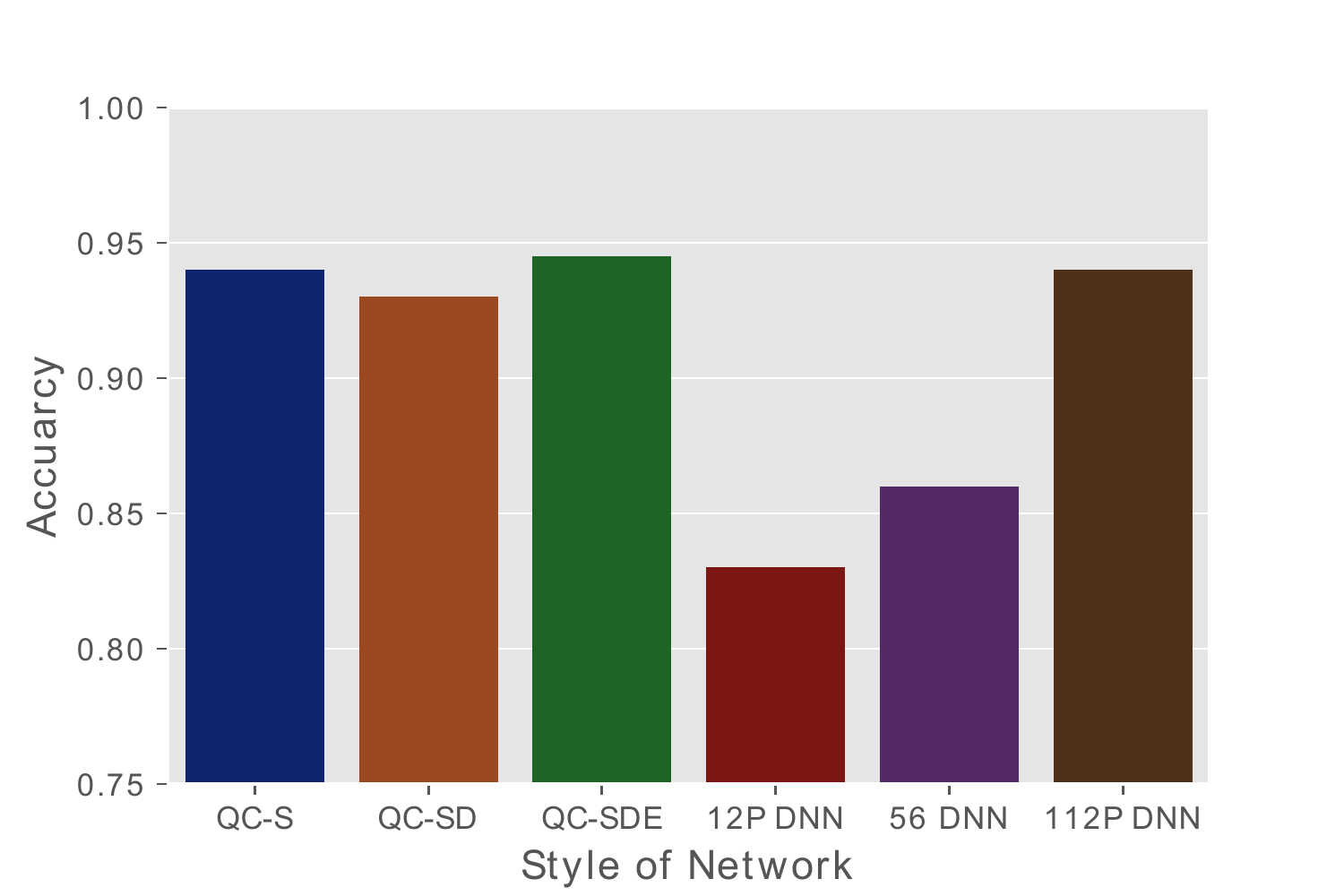}
\caption{Iris Accuracy.}
      \label{fig:acc_iris_archs}
\end{subfigure}
\begin{subfigure}{.32\textwidth}
\centering
         \includegraphics[width=1.0\linewidth]{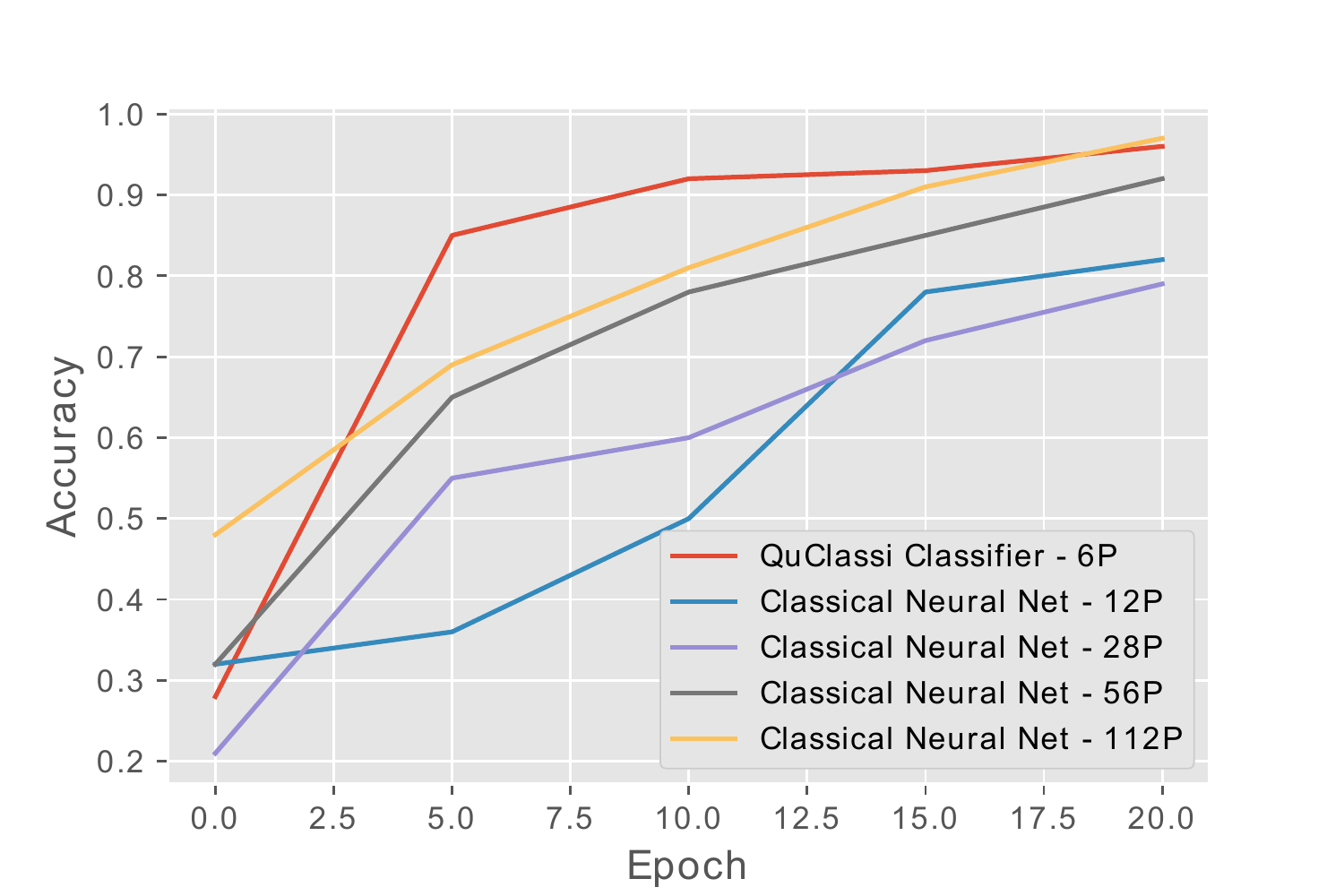}
\caption{Multiple Parameter Settings.}
      \label{fig:acplottime}
\end{subfigure}
\caption{Iris Dataset}
\end{figure*}

\subsection{System implementation and Experiment settings}

We implement \sol~ with Python 3.8 and IBM Qiskit Quantum Computing simulator package. 
The circuits were trained on a single-GPU machine (RTX 2070). 
When analyzing our results, certain important architectural terms are used. 
We describe our quantum neural networks by which qubit-manipulation layer, or what combination of layers were used. 
Discussed under Section~\ref{sec:design} with the three types of qubit manipulation, these types are a type of layer and have a respective name that lists below. On the figures, we use \texttt{QC-S, QC-D, QC-E, QC-SDE} to represent them.
\begin{itemize}
    \item \texttt{QC-S}: A layer that performs one single qubit unitary per qubit is namely a Single Qubit Unitary Layer.
    \item \texttt{QC-D}: A layer that performs a dual Qubit unitary per qubit combination is a Dual Qubit Unitary Layer.
    \item \texttt{QC-E}: A layer that entangles qubits through controlled rotations is an Entanglement Layer.
    \item \texttt{QC-SD},~\texttt{QC-SDE}: A \sol~that consists of multiple layers.
\end{itemize}


In the evaluation, we utilize Tensorflow, a neural network program framework by Google and Tensorflow Quantum, a quantum programming framework of traditional Tensorflow.

When comparing to traditional neural network methods, our primary interest is to design comparable networks with a similar number of parameters. Therefore when referencing a traditional neural network we use the term Deep Neural Network, or DNN. Furthermore, this is preceded by $k$P, where $k$ is a number indicating the total parameter count of the network. For example, DNN-12 means a deep neural network with 12 parameters. The network is trained using the same learning rate as \sol, is designed to have one hidden layer, with an output SoftMax layer. The optimizer utilized is a Stochastic Gradient Descent optimizer, the same that is used for \sol. Finally, the dataset fed to the classical neural network is the same dataset fed to the quantum neural networks, i.e. normalized data points between 0 and 1, post data-processing techniques.

Furthermore, we compare \sol~ with two quantum-based solutions: {\bf Tensorflow-Quantum (TFQ)} and {\bf QuantumFlow}~\cite{jiang2020}. Tensorflow Quatum provides its classification implementation in~\cite{tfq}, however, it only works for binary classifications. QuantumFlow, a state-of-the-art solution, works for both binary and multi-class classifications. In the comparison, we use the results from {\bf QF-pNET}, the version that obtains the best results of QuantumFlow.

\subsection{Quantum Classification: Iris Dataset}
\label{sec:iris}
A common data set that machine learning approaches are tested on for a proof of concept is the Iris dataset~\cite{iris}. This dataset comprises of 3 classes, namely Setosa, Versicolour and Virginica , and a mix of 150 data points, each containing 4 numeric data points about one of the flowers. We implement QuClassi on this data set as it provides the proof of concept for a quantum multi-class classification architecture. For this dataset to be encoded in quantum states, we perform the quantum data encoding described in~\ref{encode}. 
The data is encoded in the quantum state by performing an $RY$ gate followed by a $RZ$ gate, encoding 2 dimensions on one qubit.

\begin{figure}[!htb]
\centering
         \includegraphics[width=0.90\linewidth]{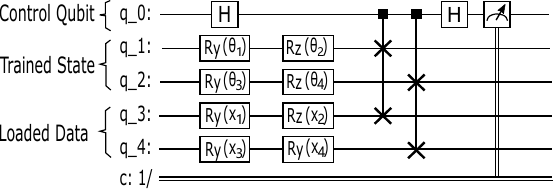}
\caption{Sample Circuit.}
      \label{fig:sampleCircuit}
\end{figure} 

To classify all three classes, there are three target qubit states $|Setosa\rangle$ (Class\_1), $|Versicolour\rangle$ (Class\_2) and $|Virginica\rangle$ (Class\_3). The dataset is separated into its respective classes, and used to train its respective quantum state. We use the methodology outlined in~\ref{training} for training these 3 quantum states. Taking the data values as $x_1,x_2,x_3\text{ and } x_4$ and trained parameters as $\theta_1,\theta_2,\theta_3\text{ and }\theta_4$ we program our quantum circuit as shown in Fig.~\ref{fig:sampleCircuit},
where the learned state is loaded into the qubits 1 and 2, and the data loaded into qubit 3 and 4. This circuit is a discriminator circuit of Class\_1, and has to be run with Class\_2 discriminator state loaded into qubits 1 and 2, and then Class\_3 with the same methodology. These probabilities are softmaxed which is then used to make a decision.
We train our circuit iteratively with a learning rate of $\alpha = 0.01$ and over 25 epochs. We illustrate our gradients, loss function and accuracy as a measure of epoch, and illustrate the significant improvement that we have attained in stability of quantum machine learning, in Fig.~\ref{fig:multi_class}.


Fig.~\ref{fig:acc_iris_archs} and \ref{fig:acplottime} plot the results of our experiments in different settings.
As can be seen, the quantum neural network architectures converge extremely quickly, and with a relatively high accuracy. Similarly parameterized classical deep neural networks perform below that of their quantum counterparts. We analyze and compare our network to classical deep neural network structures and show that the quantum neural network learns at a significantly faster rate than classical networks. We test and compare our network design of 12 parameters compared to a varying range of classical neural networks parameter counts, between 12 and 112. This is illustrated in Fig.~\ref{fig:acplottime}, where multiple classical deep neural networks of varying parameter counts were compared with the our architecture. 
Our network was able to learn to classify the three classes much quicker than any of the other neural network designs, and attained higher accuracy's at at almost all epoch iteration. 

\begin{figure}
\begin{subfigure}{.23\textwidth}
  \centering
  \includegraphics[width=.9\linewidth]{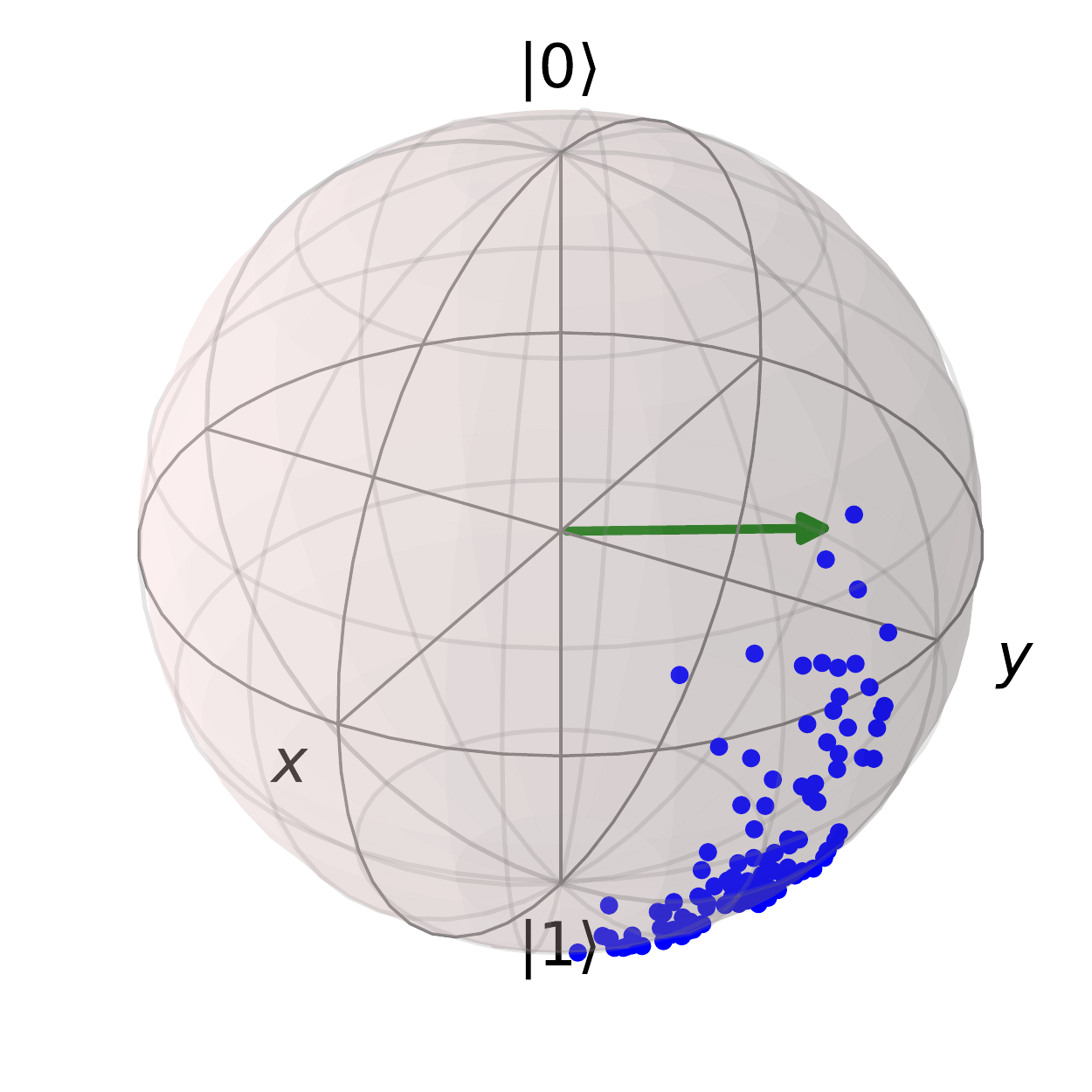}  
  \caption{Qubit 1 - 0 Epochs}
  \label{fig:sub-first}
\end{subfigure}
\begin{subfigure}{.23\textwidth}
  \centering
  \includegraphics[width=.9\linewidth]{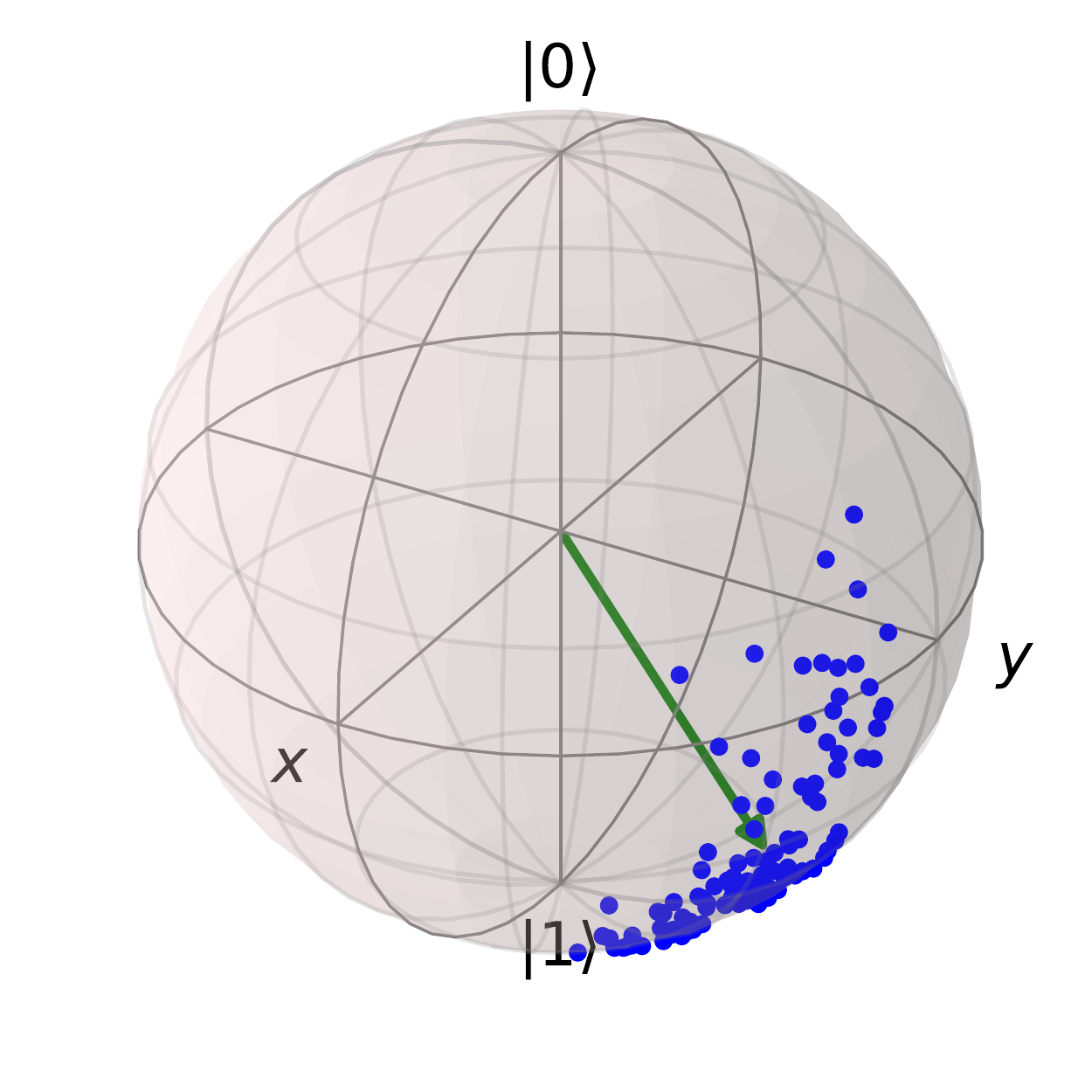}  
  \caption{Qubit 1 - 10 Epochs}
  \label{fig:sub-third}
\end{subfigure}
\caption{Identify 0 (Epoch 1 vs 10).}
\label{fig:quantum_state_evolution}
\end{figure}

\subsection{Quantum Classification: MNIST dataset}
Although the Iris dataset was able to provide a proof of concept, and the potency of this architecture, a more challenging task is classifying the MNIST dataset~\cite{mnist}. Furthermore, in comparing our architecture to others, MNIST is a common benchmark presented in literature. The MNIST is a dataset of hand written digits of resolution $28\times28$, 784 dimensions. Unfortunately, utilizing the evaluation data-encoding technique, it is impossible to conduct experiments on near-term quantum computers and simulators due to lacking of qubits and complexity of computation. 
Hence, we need to scale the dimensionality down. We make use of Principal Component Analysis (PCA~\cite{lloyd2014quantum}), as this is an efficient down scaling algorithm that has the potential to work on a quantum computer. We downscale from our 784 dimensions to 16 dimensions for quantum simulation. As for IBM-Q experiments, we make use of 4 dimensions due to the qubit-count limit of publicly available quantum computers. We make a note of that this PCA-ed data is the same data being fed to the classical neural networks. Furthermore, due to the cost-per-run model of IonQ, we do the majority of our architecture's evaluation on IBM-Q.

\sol~ allows us to perform 10-class classification on the MNIST dataset, unlike few other works that tackle binary classification problems such as (3,6), or extend into low-class count multi-class classification such as (0,3,6). 
To understand how our learning state, we visualize the training process on learning to identify 0 against a 6 by looking at the final state that is to be passed to the SWAP test. As visualized in Fig.~\ref{fig:quantum_state_evolution}, an initial random quantum state is visualized to learn to classify a 0 against a 6. It is of note that the state visualization does not encapsulate possible learned entanglements, but serves as a visual aid to the learning process.
As shown in Fig.~\ref{fig:quantum_state_evolution}, we see the evolution of the identifying state through epochs. Green arrows indicate the deep learning final state, and blue points indicate training points. We see initial identifying states to be random, but rotates and moves towards the data, such that its cost is minimized.

\begin{figure}[!htb]
\centering
         \includegraphics[width=1\linewidth]{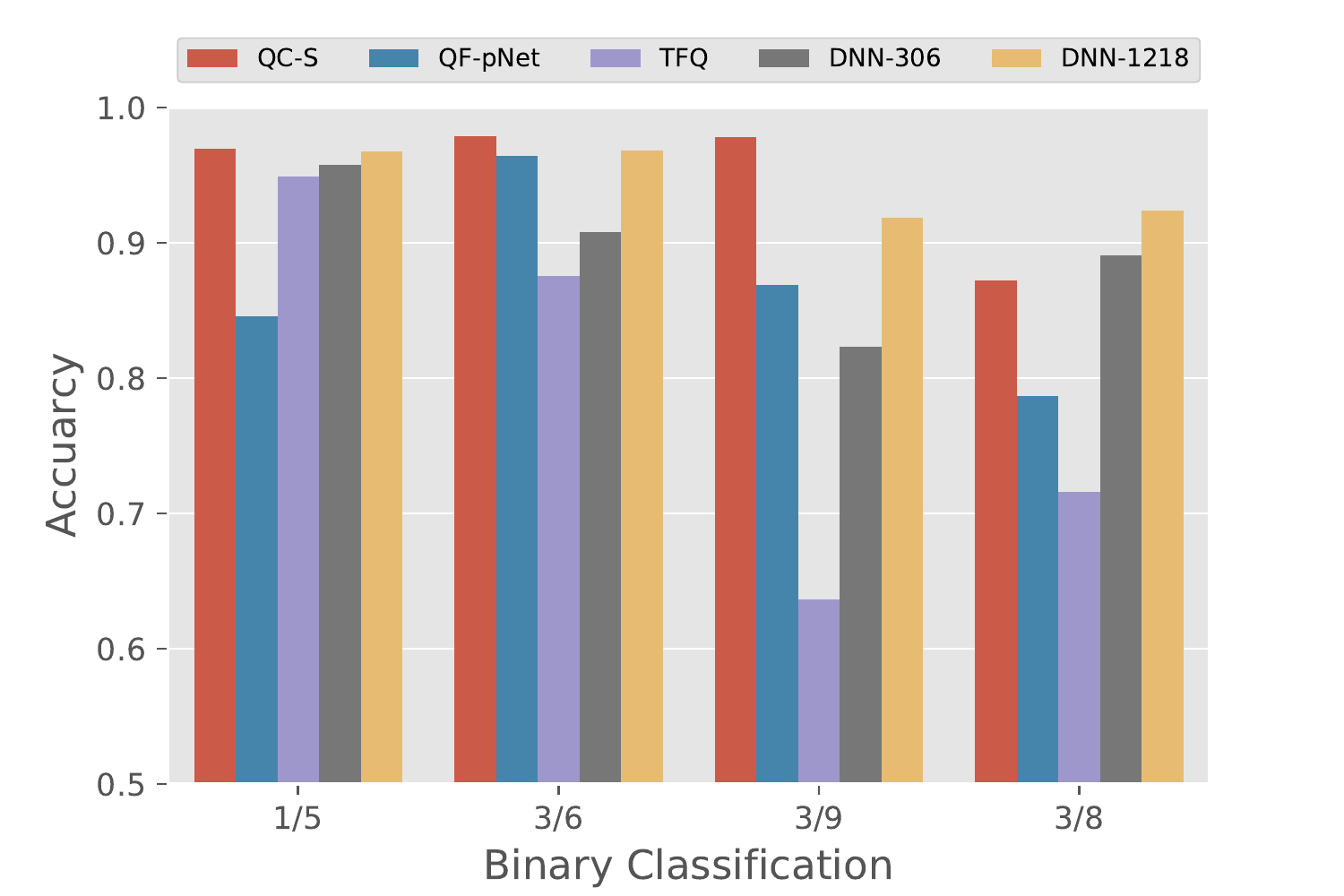}
\caption{Binary Classification Comparison.}
      \label{fig:cr}
\end{figure} 

\subsubsection{Binary Classification}
To evaluate our network architecture and its efficiency, we compare our system performance with other leading quantum neural network implementations. We compare our architecture to two leading quantum deep learning models, namely Tensorflow Quantum ({\bf TFQ})~\cite{broughton2020tensorflow} and QuantumFlow ({\bf QF-pNet})~\cite{jiang2020} and run classical deep neural networks that attain similar accuracy based on similar learning settings.  Within this section, our quantum neural network is comprised of $17$ qubits, with a total of $32$ trainable parameters in the \texttt{QuClassi-S}, the single layer setting.

The comparison of binary classification results is visualized in Fig.\ref{fig:cr}.  
Clearly, \sol~ consistently outperforms Tensorflow-Quantum, with the largest improvement of $53.75\%$ being seen in our (3,9) classification, with an accuracy of $97.80\%$. When comparing with QF-pNet, \sol~ also surpasses it, with the largest margin being attained in the (1,5) classification, where we observe a $14.69\%$ improvement over QF-pNet. \sol~ achieves $96.98\%$ and QF-pNet $84.56\%$. Therefore, we consistently perform above the binary classification tasks that shown in their works.

As for comparing with a classical deep neural network, to attain similar accuracy's, $1218$ parameters were used on similarly parameterized networks. This is in contrast to our $32$ parameters per \texttt{QuClassi-S} network. This is a substantial parameter count reduction of $97.37\%$.

\subsubsection{Multi-class Classification}
One significant contribution of our quantum neural network architecture is its multi-class adaptability. This can be relatively ambiguous in existing solutions. \sol~ provides substantially better multi-class classification accuracies.

\begin{figure}[!htb]
\centering
         \includegraphics[width=1\linewidth]{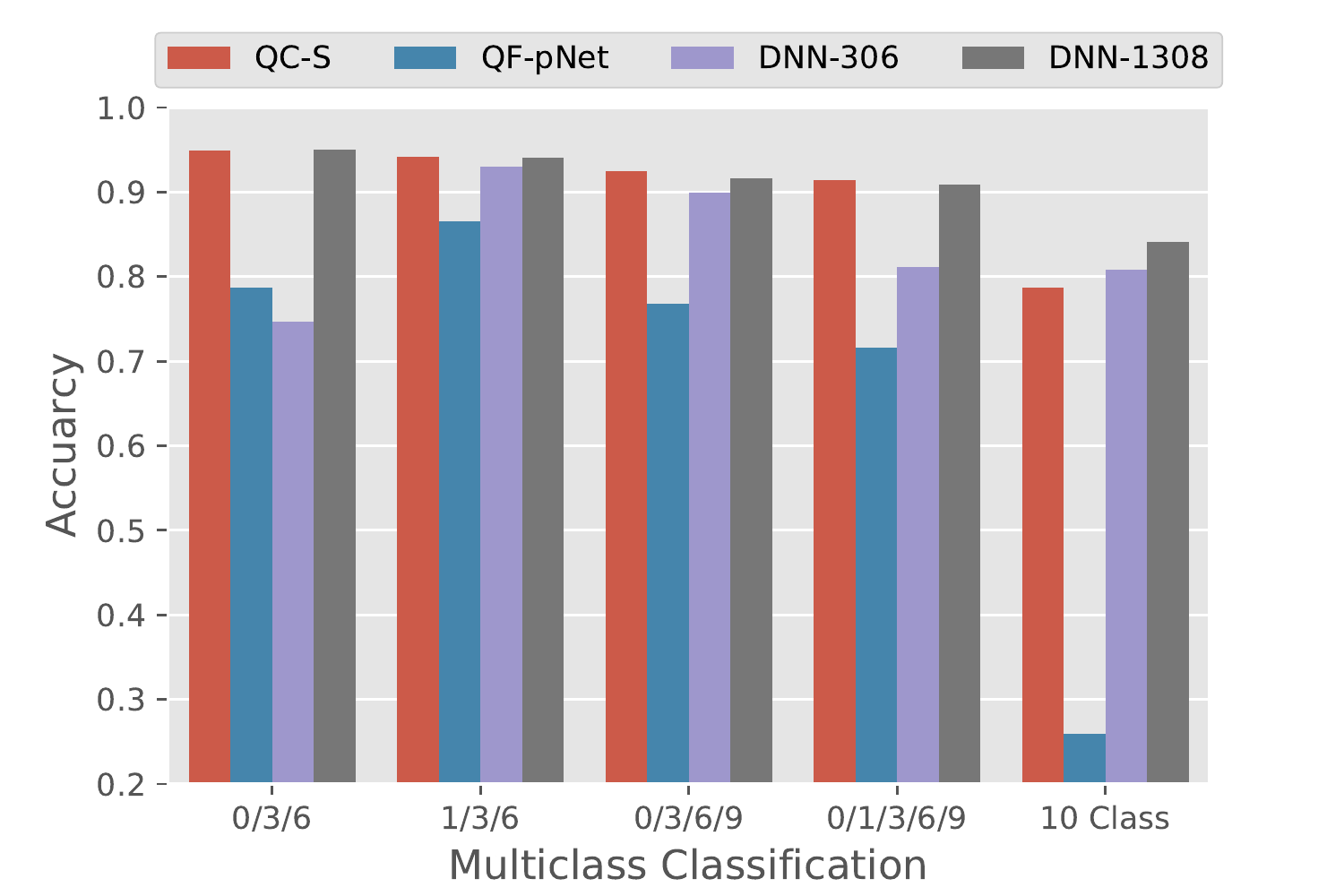}
\caption{Multi-class Classification Comparison.}
      \label{fig:mnist}
\end{figure}

With the multi-class classification results visualized in \ref{fig:mnist}, we observe that \sol~ consistent outperforming of QF-pNet in 3 class classification. For example, \sol~ achieves 94.91\% and 94.18\% for (0,3,6) and (1,3,6), comparing with 78.70\% and 86.50\% obtained by QF-pNet, which leads to accuracy increases of 20.60\% and 8.88\%. In 4 class classification, \sol~ gains $20.46\%$ (92.49\% vs 76.78\%). As the number of classes increase, \sol~ outperforms QF-pNet more with $27.72\%$ (91.40\% vs 71.56\%) in 5-class and $203.00\%$ (78.69\% vs 25.97\%) 10-class classification. In QuantumFlow (QF-pNet), most of the training is done on the classical computer, where the traditional loss function is in use. With \sol, however, we employ a quantum-state based cost function that can fully utilize the qubits.

A trend emerges from these improvements, highlighting how \sol~ performs on higher class count.  In comparing to classical deep neural networks that can achieve similar accuracies, \sol~ attains a $96.33\%$ reduction in parameters on 5-class classification (48 vs 1308), and a $47.71\%$ reduction in parameters on 10-class classification (160 vs 306) in the quantum setting.

\begin{figure}[!htb]
\centering
         \includegraphics[width=1\linewidth]{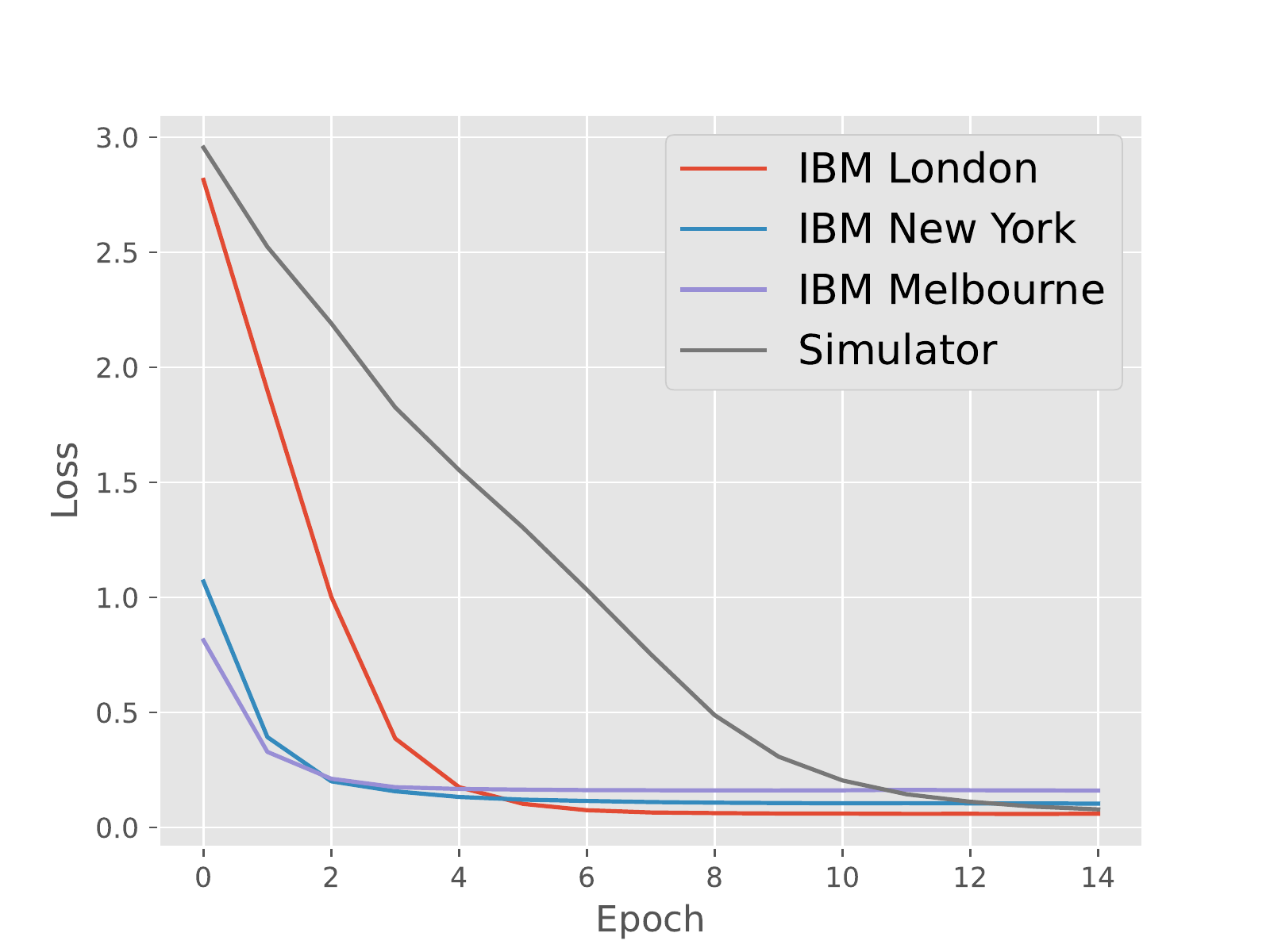}
\caption{Quantum Computer Training: Iris.}
      \label{fig:rqct}
\end{figure}

\subsection{Experiments on IBM-Q}

As a proof of concept, we evaluate our result on actual quantum computers. 
For Iris dataset (4 dimensions), the \texttt{QuClassi-S} utilize 5 qubits. 
With MNIST dataset, however, the previous simulations were ran with a total of 17 qubits. It is relatively difficult to perform the same experiments on publicly quantum computers due to limited qubits available. Therefore, we downscale our data to $4$ dimensions with PCA, such that we can make use of 5-qubit IBM-Quantum Computers. For experiments of Iris dataset, we utilize multiple IBM-Q sites around the world. For MNIST dataset, the experiment is conducted at IBM-Q Rome.  

Fig.~\ref{fig:rqct} presents the results of Iris dataset.
Despite the fact IBM-Q quantum computer's communication channel introduces an extra overhead and shared public interfaces leading to large backlogs and queues, we attain a result of IBM-Q London site with an accuracy of 96.15\% on Iris dataset, which is similar to the value obtained by simulations. 
We train each epoch through Iris dataset with 8000 shots (number of repetitions of each circuit) to calculate the loss of the circuit. Based on our observation and previous research in this area \cite{8914134,8491885} the integrity of physical qubits and \texttt{T1}, \texttt{T2} errors of IBM-Q machines could vary~\cite{tannu2019not}, however, our design managed to attain a solution after few iterations, comparable to the simulator results we attain. 
Running experiments on actual quantum computers generated stable results and accuracy similar to the simulators. As seen in Fig \ref{fig:rqct}, the loss of the quantum circuit converges similarly on a real quantum circuit, in different IBM-Q sites, to that of a simulator.

\begin{figure}[!htb]
\centering
         \includegraphics[width=1\linewidth]{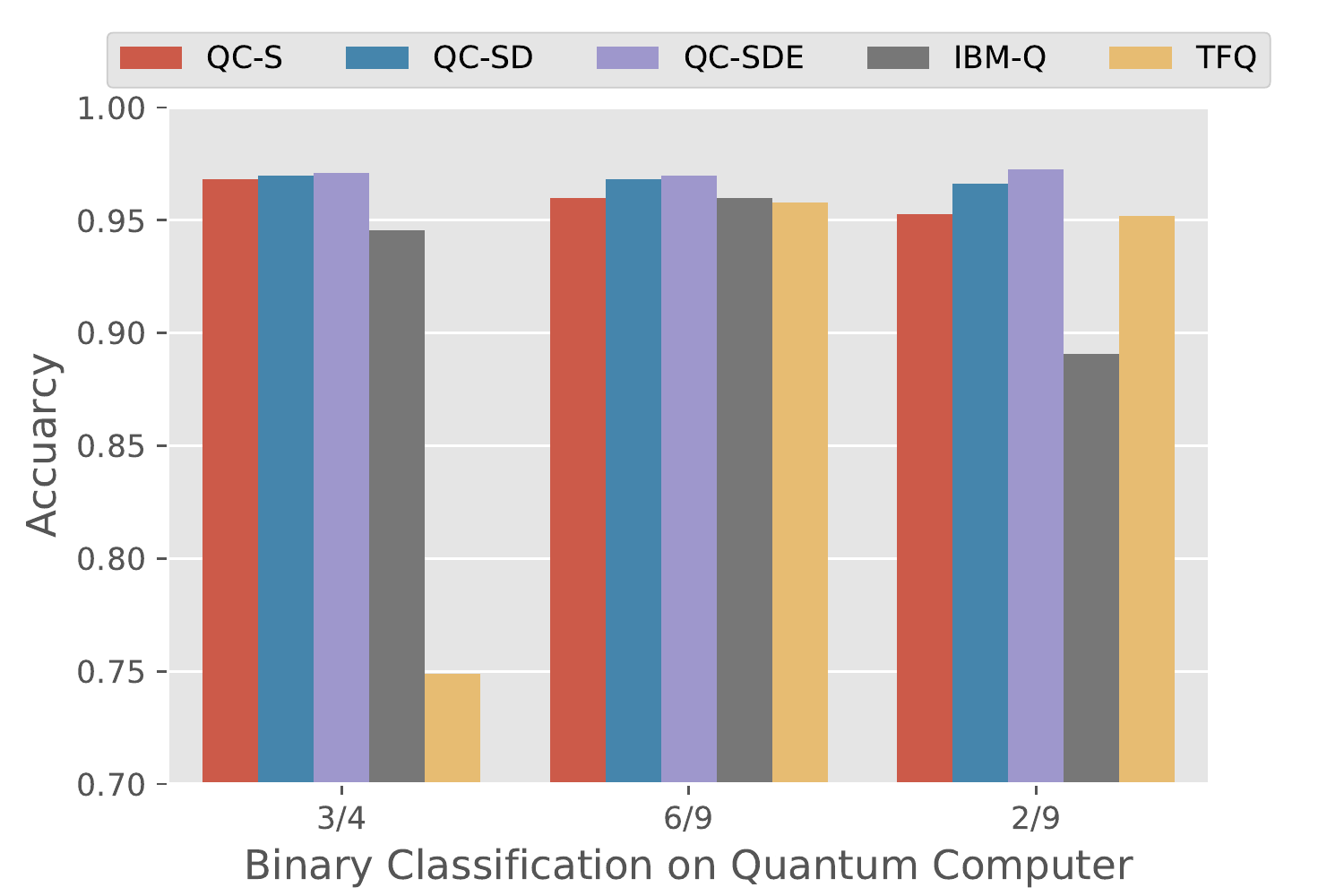}
\caption{Quantum Computer Accuracies: MNIST.}
      \label{fig:rqc_acc}
\end{figure}

Fig.~\ref{fig:rqc_acc} plots the results of \texttt{QuClassi-S} with 4-dimension MNIST dataset on IBM-Q Rome site. We compare the real quantum experiments with simulations of TFQ and three versions of \sol. With (3,4) and (6,9), the experiments achieve similar results as simulations. For example, the measured difference of the experiment on the IBM-Q quantum computer and simulation of (6,9) is 0.2\% (95.98\% 96.17\%). With (3,4), the accuracies of QC-S are 94.56\% and 96.81\% for real quantum computer and simulation, respectively. In the experiment of (2,9), however, the measured difference is 6.9\%, where the experiment result is 89.09\% and simulation, 95.27\%. The larger difference is due to noise on quantum processors that depends on the processors performance and topology. We further evaluate \sol on IonQ's trapped ion quantum processor. We test (3,6) on IonQ and IBM-Q's Cairo machine. An ideal accuracy of 97.80\% is attained, with IonQ attaining 80.00\% and IBM-Q Cairo 72.00\%. This difference can be attributed to the fully-connected nature of trapped IonQ, allowing for 0 SWAP operations, compared to IBM-Q Cairo, requiring 21 CNOT operations due to topological constraints, attributing to this discrepancy in accuracy. 
In addition, we can see that three versions of \sol~ perform similarly and consistently. This is the same trend as we have seen in Section~\ref{sec:iris} that, with low-dimensional data, the improvement of a deeper network is limited.

\section{Discussion and Conclusion}
\label{sec:conclusion}
In this project, we propose \sol, a novel quantum-classic architecture for multi-class classification. The current accuracy attained by \sol~ performing binary classification on the MNIST data set is competitive, beating out  Tensorflow Qauntum by up to $53.75\%$ on binary classification. \sol~also outperforms QuantumFlow by up to $14.69\%$ on binary classification, and up to $203.00\%$ on multi-class classification. Furthermore, to the best of our knowledge, \sol~ is the first solution that tackles a 10-class classification problem in the quantum setting with a performant result. 
Additionally, comparing \sol~ with similarly performed classical deep neural networks, \sol~ outperforms them by learning significantly faster, and requiring drastically less parameters by up to $97.37\%$ on binary classification and up to $96.33\%$ in multiclass classification.


Our work provides a general step forward in the quantum deep learning domain. There is, however, still significant progress to be made. Performing the 10-class classification of MNIST lead to a $78.69\%$ accuracy, which is relatively poor when compared to its classical counterparts. Although using many more parameters, classical counterparts are able to reach an accuracy of near 100\%, \sol~ shows the potential that quamtum may bring to us. Our future work will focus on improving the multi-class classification that aims to further improve the accuracy. Moreover, understanding the low-qubit representation of quantum data and its implications within the quantum based learning field should be investigated.

\section{Acknowledgements}
\label{sec:acknowledgements}
This material is based upon work supported by the U.S. Department of Energy, Office of Science, National Quantum Information Science Research Centers, Co-design Center for Quantum Advantage (C2QA) under contract number DE-SC0012704. We acknowledge support from Microsoft's Azure Quantum \cite{Azure} for providing credits and access to the ion-trap quantum hardware used in this paper. The Pacific Northwest National Laboratory is operated by Battelle for the U.S. Department of Energy under contract DE-AC05-76RL01830.

\bibliography{references}
\bibliographystyle{mlsys2022}


\clearpage

\appendix

\section{Artifact Description}
\label{sec:artifact}
\subsection{Abstract}

The following artifact appendix contains information used to reproduce the QuClassi results in the described paper. The evaluation is done on a quantum computing simulator, and can be connected to a IBM-Q device using IBM-Q Backend. Training the network is computationally expensive, hence changing the subsample variable will change the amount of data being fed to the network. On induction, the network will train, the variables will be saved, and the performance of the network will be posted. 

\subsection{Artifact check-list (meta-information)}

{\small
\begin{itemize}

  \item {\bf Data set:} MNIST, Iris
  \item {\bf Run-time environment:} Python 3.7, Qiskit
  \item {\bf Hardware: } CPU, GPU for simulations, IBM-Q for real quantum experiments
  \item {\bf Execution}: Single file execution, main.py
  \item {\bf Metrics:} Accuracy 
  \item {\bf Output:}  Accuracy
  \item {\bf How much disk space required (approximately)?:} Approx. 10mb 
  \item {\bf How much time is needed to prepare workflow (approximately)?:} Time to install dependencies
  \item {\bf How much time is needed to complete experiments (approximately)?:} Up to 7 days for full dataset 
  \item {\bf Publicly available?:} Yes
\end{itemize}

\subsection{Description}

\subsubsection{How delivered}

To access the code, see \url{https://github.com/Samuelstein1224/QuClassiExample}

\subsubsection{Hardware dependencies}

Quantum Processor access for real world validation, CPU for simulator simulation 

\subsubsection{Software dependencies}
Python 3.7.0 with packages Qiskit,Tensorflow,Numpy and sklearn

\subsubsection{Data sets}
MNIST and Iris dataset.

}
\subsection{Installation}
Installation of Python, followed by using pip install suffices for all packages (Qiskit, Tensorflow, Numpy, Matplotlib and Sklearn)

\subsection{Evaluation and expected result}

On induction, the model will train a quantum neural network over the MNIST dataset, with SUBSAMPLE data points. The computational cost of training can be high, therefore changing this to what best suits the system you are running on is best. Accuracies expected are in the high 90's (approx. 96\% accuracy), with some very slight variance due to the random nature of quantum state sampling.
\subsection{Experiment customization}

To change the simulator, the backend must be changed within the code. This can be used to apply noise etc. For the system to be run on a real quantum processor, the backend must be changed from a simulator to a real quantum processor. 
\subsection{Methodology}

Submission, reviewing and badging methodology:

\begin{itemize}
  \item \url{http://cTuning.org/ae/submission-20190109.html}
  \item \url{http://cTuning.org/ae/reviewing-20190109.html}
  \item \url{https://www.acm.org/publications/policies/artifact-review-badging}
\end{itemize}

\end{document}